\documentclass[12pt]{article}\usepackage[]{graphicx}\usepackage[]{color}
\makeatletter
\def\maxwidth{ %
  \ifdim\Gin@nat@width>\linewidth
    \linewidth
  \else
    \Gin@nat@width
  \fi
}
\makeatother

\definecolor{fgcolor}{rgb}{0.345, 0.345, 0.345}

\usepackage{framed}
\makeatletter
 {\par\unskip\endMakeFramed%
 \at@end@of@kframe}
\makeatother

\definecolor{shadecolor}{rgb}{.97, .97, .97}
\definecolor{messagecolor}{rgb}{0, 0, 0}
\definecolor{warningcolor}{rgb}{1, 0, 1}
\definecolor{errorcolor}{rgb}{1, 0, 0}
\newenvironment{knitrout}{}{} % an empty environment to be redefined in TeX

\usepackage{alltt}

\usepackage{amssymb,amsmath,amsthm}
\usepackage{graphicx}
\usepackage{geometry}
\usepackage{natbib}
\usepackage{hyperref}

\newcommand{\Exp}[1]{{\text{E}}[ #1 ]}

\title{Coverage Properties of Empirical Bayes Intervals}
\author{
Peter Hoff \\ Department of Statistical Science \\ Duke University}
\IfFileExists{upquote.sty}{\usepackage{upquote}}{}
\begin{document}
\maketitle

\begin{abstract}
This note is an invited discussion of the article 
 ``Confidence Intervals for Nonparametric Empirical Bayes Analysis'' by Ignatiadis and Wager. In this discussion, I review some goals of empirical Bayes 
data analysis and the contribution of 
Ignatiadis and Wager. Differences between across-group inference and 
group-specific inference are discussed. Standard empirical Bayes interval 
procedures focus on controlling the across-group average coverage rate.
However, if group-specific inferences are of primary interest, 
confidence intervals with group-specific coverage control may be preferable. 

\end{abstract}

\section{Introduction}
Empirical Bayes methods are commonly used in the analysis of data coming from 
multiple related populations or groups, where often the goal is to 
obtain a parameter estimate for each individual group. 
Unlike a ``direct estimate'' that uses data only from a
given group to construct that group's estimate, 
an empirical Bayes estimate for a given group may use data from all of the 
groups. As a result, empirical Bayes estimators can have lower variance, 
and typically lower risk, than direct estimators, at least 
on average across the groups. 

The optimal amount of across-group information sharing is 
determined by the actual across-group heterogeneity, which is 
generally unknown. Empirical Bayes methods based on 
plug-in empirical estimates
of this heterogeneity can in some cases provide asymptotically 
optimal group-level estimators. 
However, as \citet{ignatiadis_wager_2021} (IW) point out, 
for practical finite-sample data analysis,
the estimated and true across-group heterogeneity 
may be quite different, and 
so proceeding with a plug-in estimate without considering a 
range of other plausible values may  result in 
misleading or incomplete inferences. 
To remedy this situation, IW provide several 
practical methods 
that more completely describe the uncertainty in 
empirical Bayes estimates.
% by accounting for the variability in 
%the estimate of 
%across-group heterogeneity. 
Specifically, they provide 
tools for constructing 
asymptotically correct frequentist confidence intervals for 
empirical Bayes estimands in some generic settings, as well as some methods for 
specific cases. 

Ignatiadis and Wager focus on confidence interval procedures 
for 
%point estimates of 
functions of the across-group heterogeneity. 
As they point out in a footnote, their focus \emph{is not} 
%confidence interval procedures of this type  \emph{do not} include
on intervals for the parameter of any specific group. However, 
in some applications
it is this latter type of interval 
that is of interest. In my comments that follow, I first consider 
how the methods developed by IW might be used to construct 
empirical Bayes posterior intervals for group-specific parameters that attain a target 
coverage rate on average across groups, which is a type of coverage rate control 
that is typical of empirical Bayes interval procedures studied in the literature. 
Such procedures fail to control group-specific frequentist coverage. For 
example, a nominal 95\%  empirical Bayes posterior interval 
for the mean of a given group 
may have a frequentist coverage rate that is arbitrarily close to zero, depending on how far 
the true mean for that group is from the means of the other groups.  
Yet in some applications, it is these outlying groups about which we are most 
concerned. For situations in which 
group-specific inferences are of primary interest, 
I describe an empirical Bayes confidence interval procedure
that maintains exact group-specific coverage, while improving upon ``direct'' 
intervals in terms of across-group average precision.
The coverage rate for this procedure is nonparametric, in the sense that 
it does not depend on a correct specification of the across-group heterogeneity.

\section{Procedures with across-group coverage control}  
We first review the model considered by IW:
Data $Z_1,\ldots, Z_n$ are to be independently randomly sampled from 
$n$ different groups, with $Z_i$ being the random variable 
to be sampled from group $i$.
Further suppose that the distribution of $Z_i$ 
has a density $p(z|\mu_i)$ for 
some $\mu_i \in \mathcal M \subset \mathbb R^p$. 
Examples presented in IW include where 
$Z_i$ is the number of insurance claims made by the $i$th insurance holder 
and $\mu_i$ is their long-term rate of making claims, 
and where $Z_i$ is the number of test questions that are correctly answered 
by student $i$ and $\mu_i$ is the probability that they will correctly 
answer an individual question.
In both of these examples the ``groups'' are individual people, and
systematic heterogeneity among the people in terms of the measured variable
is quantified by 
heterogeneity among $\mu_1,\ldots,\mu_n$. 
A somewhat different type of example, often encountered in the 
small-area estimation literature, is 
where each participant in a large survey falls into one of $n$ categories, and 
$Z_i$ is the sample mean for the survey participants falling into category $i$. 
For example, in Section 4 we consider data from 
a survey of household radon levels in Minnesota.  
In this application, 
$\mu_i$ is the mean 
household radon level in 
county $i$, and  $Z_i$ is the sample mean of observations
from the survey that are in county $i$. 
Nearly all of the counties 
in the state 
are represented in the dataset, and the goal is to infer 
$\mu_i$ for each of these counties. 
%, and so the groups in this case are not 
%a random sample. 

%Just as Bayes procedures are concerned with statistical 
%performance on average across possible values of a single parameter  
%(i.e.\ the Bayes risk), the study of 
%empirical Bayes procedures has typically focused 
%on performance on average across groups, 
%distribution of the $\mu_i$'s \citep{jiang_zhang_2009}. 

Empirical Bayes procedures are often motivated by imagining 
(justifiably or not) that 
the $n$ groups that appear in the dataset
are an i.i.d.\ sample from some larger population of groups, 
and therefore $\mu_1,\ldots,\mu_n$ is an i.i.d.\ sample from 
some distribution $G$. 
If $G$ were known, then 
upon observing $Z_i=z_i$
one could compute the posterior mean estimator 
of $\mu_i$ as $\theta_G(z_i)$ where 
$\theta_G(z) = \Exp{ \mu | Z = z }$. 
This estimator is optimal in terms of  
prior (or marginal) 
 expected squared-error 
loss, on average over both $\mu_i$ and $Z_i$. Specifically, 
$\theta_G(z)$ minimizes the 
Bayes risk 
\begin{equation}
 R_G(\theta) =   \int  \int  ( \mu - \theta(z) )^2 \,  p(z|\mu) g(\mu) \, dz \, d\mu, 
\end{equation}
where $g$ is the density of $G$. 
Since $G$ is generally unknown, standard practice is to construct 
an estimate $\hat G$ from $z_1,\ldots, z_n$, and then use it in place 
of $G$ when computing the posterior mean estimator, resulting in the 
empirical Bayes estimator 
$\theta_{\hat G}$. 

From IW's perspective,  
$\theta_G(z)$ is a an estimand, and $\theta_{\hat G}(z)$ is 
an estimate. If a data analysis 
includes $\theta_{\hat G}(z)$, it should also include
some description of other plausible values of 
$\theta_{G}(z)$, such as those provided by a confidence interval. 
This perspective is well-motivated by their insurance claim example:
A common insurance premium $c_z$ in the next year
will be applied to all individuals who make $z$ claims this year. 
The appropriate value of $c_z$ 
 is determined by the expected number of claims made next year 
  by one of these people, 
which under the model is 
$\Exp{ \mu | Z= z}$.
Clearly, a confidence interval for $\Exp{ \mu | Z= z}$ is of use 
in this application. More broadly, when the targets of inference 
involve averages or expectations across 
 different $\mu$-values of 
multiple individuals or groups, 
then the procedures provided by IW are a welcome 
contribution to  empirical Bayes methodology. 
 
In other applications the individual $\mu_i$'s are the targets of inference,
in which case a confidence interval for  $\mu_i$ is needed, instead of one for 
$\Exp{ \mu | Z= z}$.
This is conceivably the case for IW's 
psychometric test example, where $Z_i$ is the number of questions out of 20
that student $i$ 
answers correctly on a standardized test. The model in this example 
is that $Z_i|\mu_i \sim$ binomial(20, $\mu_i$), and so 
$\mu_i$ represents the test-taking ability of student $i$. 
Upon observing $Z_i=z$, estimating $\mu_i$ 
with an empirical Bayes version of $\Exp{\mu | Z=z}$ is 
very reasonable. However, a confidence interval 
for $\Exp{\mu | Z=z}$ is not a confidence interval for $\mu_i$. 
In particular, 
$\Exp{\mu | Z=z}$ is an average of $\mu$-values across the set of 
students who obtain a score of $z$, and so a confidence 
interval for $\Exp{ \mu |Z=z}$ is a confidence interval for 
this across-student average ability, rather than a confidence interval for the 
single $\mu$-value of a specific student. Furthermore, 
as the number of students $n$ in the study increases to infinity, 
the width of a 
confidence interval for 
$\Exp{ \mu |Z=z}$ should decrease to zero, whereas the width of
a confidence interval 
for $\mu_i$ will not decrease (without bound)
unless the number of questions 
answered by student $i$ increases.

Ignatiadis and Wager point out this distinction as a footnote in their 
Introduction, and clarify that the focus of their article is on 
confidence intervals for across-group estimands, that is, functions 
of $G$ and not on individual $\mu_i$'s. 
However, I speculate that their methods, or an extension thereof, 
may also be used to make 
a standard type of ``empirical Bayes'' confidence interval for each $\mu_i$.  
These intervals are generally constructed as follows: If $G$ were known,  
then for each $z$ we could compute a quantile-based posterior interval
$C(z)= (l(z),u(z))$ such that
%of the posterior density of $\mu$ given $Z=z$ such that 
\begin{equation}
 \Pr(\mu\in C(z)| Z=z) = \Pr(  l(z) < \mu  < u(z) | Z=z ) = 1-\alpha. 
\end{equation}
Note that since the interval has $1-\alpha$ coverage 
conditionally for each $z$, it also has $1-\alpha$ coverage 
with respect to the joint distribution of $\mu$ and $Z$, and 
so in this sense, the \emph{average} frequentist coverage rate across groups 
(average across values of $\mu$ with respect to $G$) is $1-\alpha$. 
%However, as will be discussed further in the next section, 
%such an interval generally does not have $1-\alpha$ coverage 
%``conditional'' on $\mu_i$. In other words, such an interval is 
%not a $1-\alpha$ ``frequentist'' interval for the fixed parameter $\mu_i$. 
%So in the remainder of this note, we will say that such intervals 
%have across-group coverage control, as opposed to having group-specific 
%coverage control. 
Since $G$ is unknown, common practice has been to replace $l$ and 
$u$ with the corresponding quantiles of an estimate $\hat G$ of $G$. 
Although typically ignored in applied practice,  
it has long been known that replacing $G$ with an estimate $\hat G$ 
affects the across-group coverage rate. To remedy this, 
%In the case that $G$ is an unknown normal distribution, 
 \citet{morris_1983} suggests widening the interval to make it 
resemble an interval from a ``full'' Bayesian posterior distribution, while
 \citet{laird_louis_1987} 
propose a bootstrap procedure to account for uncertainty in $G$. 
To see how IW's methods
might provide an alternative approach, 
first consider the  task 
of making a one-sided upper confidence 
bound $u(z)$ for an unknown $\mu$-value, having the property 
that 
\begin{equation}
  \Pr(  \mu > u(z) | Z=z )  \leq \alpha. 
\end{equation}
Such a procedure can be constructed using 
the ideas of IW, by forming a confidence 
region for a specific functional of $G$, 
the quantile function of the conditional distribution of 
$\mu$ given $Z=z$. For $\alpha_1<\alpha$,  
let  $v(z)$ satisfy 
\begin{equation} 
\Pr( \mu >  v(z) | Z=z ) = \alpha_1. 
\end{equation}
Since $G$ is unknown, so is $v(z)$, but suppose we can obtain 
an $\alpha-\alpha_1$ upper confidence bound $u(z)$ for $v(z)$, so that 
\begin{equation}
 \Pr( v(z) > u(z) ) \leq  \alpha-\alpha_1. 
\end{equation}
Then we have 
\begin{align}
\Pr( \mu > u(z) |Z=z  )  & =   
 \Pr( \mu > u(z) , \mu > v(z)|Z=z) + 
 \Pr( \mu > u(z) , \mu< v(z)|Z=z)  \\ 
% \Pr( v(z)>  \mu > u(z)|Z=z)  \\
& \leq  
 \Pr( \mu > v(z)|Z=z) + 
 \Pr(  v(z)  >  u(z))   \\ 
 & \leq \alpha_1 + ( \alpha - \alpha_1 ) =\alpha. 
\end{align}
To clarify the way in which $u(Z_i)$ is a one-sided confidence region 
for $\mu_i$, we have 
\begin{align}
 \Pr( \mu_i < u(Z_i) ) &= 
  \int \int  1( \mu  <  u(z) ) \, p(z|\mu)dz  \, g(\mu)d\mu   \\
  &= \int \int 1( \mu  <  u(z) ) \,  g(\mu|z)d\mu \,    p(z)dz   \\ 
&=  \int  \Pr( \mu< u(z) | Z=z )\,  p(z) dz  \\
&\geq  \int  (1-\alpha) \, p(z) dz = 1-\alpha, 
\end{align} 
where $g(\mu|z)$ is the conditional density of $\mu$ given $Z=z$, and 
$p(z)$ is the marginal density of $Z$. 
A lower confidence bound $l(z)$ 
 for $\mu_i$ may be similarly constructed, and then combined with 
$u(z)$ to form a confidence interval $C(z)=(l(z),u(z))$ with the property 
that 
$\Pr(  \mu_i\in C(Z_i) )\geq 1-\alpha$. 
%If IW's approach can do this without making parameteric assumptions 
%about $G$, then its applicability would go beyond that of 
%  \citet{morris_1983}'s approach for the hierarchical normal model, 
%and perhaps make it similar to \citet{laird_louis_1987}'s type II 
%bootstrap method, which treats $G$ nonparametrically and 
%the conditional 
%distribution of $Z$ given $\mu$ parameterically. 

\section{Lack of group-specific coverage control} 
As mentioned above, a $1-\alpha$ empirical Bayes posterior interval
$C(z_i) = (l(z_i),u(z_i))$  
for $\mu_i$ given $Z_i=z_i$ is 
typically constructed to have approximate $1-\alpha$
\emph{posterior coverage}
\begin{equation}
  \Pr( \mu_i \in C(z_i)  | Z_i= z_i)  \equiv 
 \int_{C(z_i)} g(\mu | z_i) \, d\mu  \approx 1-\alpha, 
\end{equation}
for every possible value of $z_i$. 
Because the conditional coverage is 
(approximately) $1-\alpha$  for every value of $z_i$, the interval also has
$1-\alpha$ \emph{marginal coverage}, meaning that 
 $ \Pr( \mu_i \in C(Z_i) )\approx 1-\alpha$, where 
this probability is with respect to the 
joint distribution of $\mu_i$ and $Z_i$. 
This in turn constrains the 
\emph{frequentist coverage} $\Pr( \mu_i \in C(Z_i) | \mu_i)$ 
of  $C(z)$ as a confidence interval procedure for a fixed but unknown $\mu_i$:
\begin{align}
1-\alpha \approx \Pr( \mu_i \in C(Z_i))  &  = 
\int  \Pr(\mu \in C(Z)| \mu)  \,  g(\mu )\,  d\mu .  
\end{align}
So by virtue of its (approximate) posterior $1-\alpha$ coverage, an empirical Bayes procedure 
must have frequentist coverage that is 
$1-\alpha$ on average across values of $\mu$ with respect to $G$. 
In practice, this means that if the $\hat G$ used to construct 
the group-specific intervals is close to the empirical distribution of $\mu_1,\ldots, \mu_n$, 
we should have 
\begin{equation}
  \sum_{i=1}^n \Pr( \mu_i \in C(Z_i)| \mu_i )/n \approx 1-\alpha. 
\end{equation}
So even in the case that the groups, and so the $\mu_i$'s, are not randomly 
sampled, we expect the across-group average frequentist coverage rate 
of the empirical Bayes intervals to be approximately $1-\alpha$. 
However, 
the frequentist coverage rate for any particular 
group could be terrible. This should be apparent from the fact that 
the difference between Bayes and 
empirical Bayes intervals 
is just that the latter use a plug-in estimate for the 
prior distribution. For any Bayesian interval procedure
 - empirical or not -
 if the value 
of $\mu_i$ is far from the center of mass of the putative 
prior distribution, the frequentist coverage 
could be arbitrarily close to zero. 
For example, suppose
we have $n$ different groups corresponding to $n$ normal populations 
with means $\mu_1,\ldots, \mu_n$ and a common known variance $\sigma^2$. 
An observation $Z_i\sim N(\mu_i,\sigma^2)$ a will be sampled 
independently for $i=1,\ldots,n$, and we wish to make a confidence 
interval for each $\mu_i$. 
The usual frequentist interval is 
\begin{equation}
 C_{U}(z_i) =\left ( z_i+  \Phi^{-1}(\alpha/2)\sigma, z_i+  \Phi^{-1}(1-\alpha/2)\sigma \right ),
\end{equation}
which is the uniformly most accurate unbiased interval (UMAU), 
and can be derived by inverting a collection of 
uniformly most powerful unbiased (UMPU) tests. Clearly, this interval 
has exact $1-\alpha$ frequentist coverage for each group $i$, no matter 
what the value of each $\mu_i$ is. We say this interval procedure 
has \emph{constant coverage}, in the sense that  
$\Pr(\mu_i \in C_{U}(Z_i)|\mu_i)$ 
is constant as a function of $\mu_i \in \mathbb R$.
A Bayes or empirical Bayes confidence interval that assumes 
$\mu_1,\ldots, \mu_n$ are an i.i.d.\ sample from a $N(\phi,\tau^2)$ population
has the form 
\begin{equation}
C_B(z_i) = \left 
( \hat\mu_i + \Phi^{-1}(\alpha/2)/\sqrt{1/\tau^2 + 1/\sigma^2 } , 
\hat\mu_i + \Phi^{-1}(1-\alpha/2)/\sqrt{1/\tau^2 + 1/\sigma^2 }\right )
\end{equation}
where
$\hat\mu_i=  ( \phi/\tau^2 +  z_i/\sigma^2 )/ (1/\tau^2+1/\sigma^2)$ 
is the  posterior mean estimator, and 
$(\phi,\tau^2)$ are the mean  and variance of the imagined normal distribution from which the $\mu_i$'s were sampled, or empirical Bayes estimates of these quantities. 
The interval width is the same for all groups, 
but each is centered around a Bayes (or empirical Bayes) estimator 
whose bias varies across groups, 
which implies that the coverage rate will vary across groups as well.

\begin{figure} 
\begin{knitrout}
\definecolor{shadecolor}{rgb}{0.969, 0.969, 0.969}\color{fgcolor}

{\centering \includegraphics[width=5in]{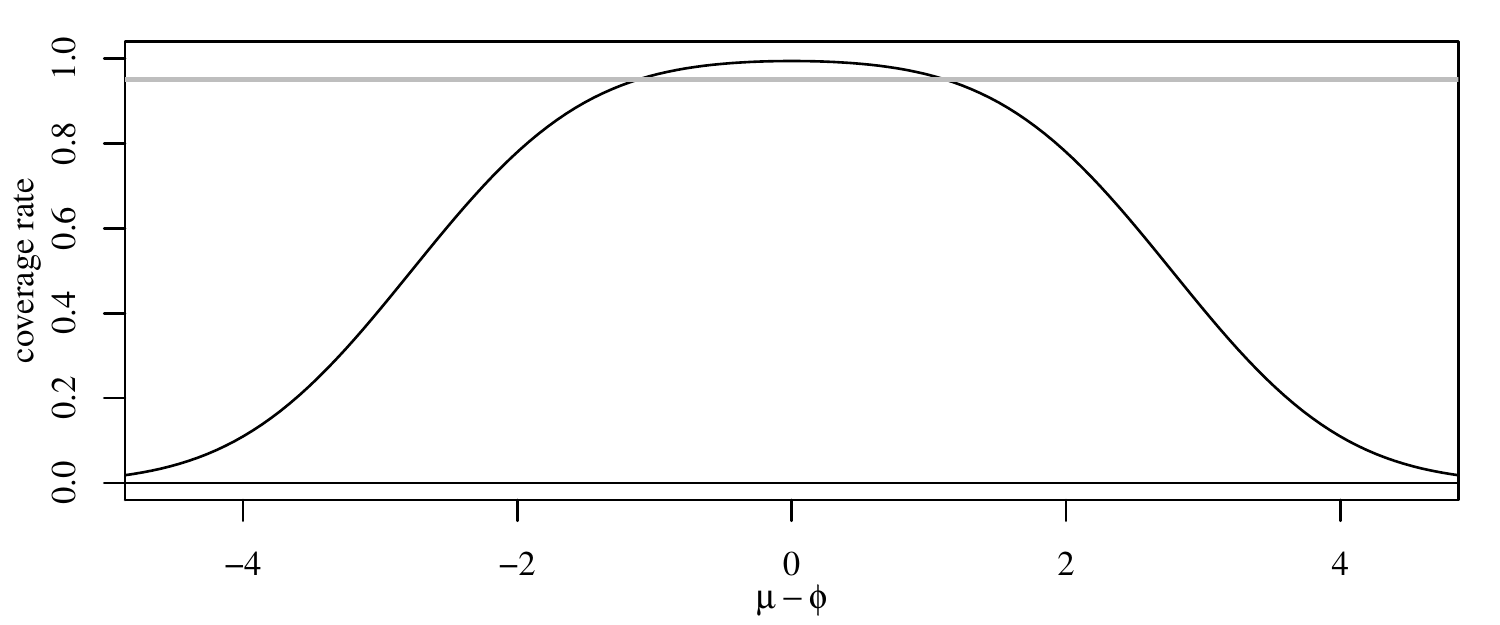} 

}

\end{knitrout}
\caption{Frequentist coverage rate of the 95\%  posterior interval,
         for $\sigma^2 = \tau^2=1$. }
\label{fig:ebcover} 
\end{figure}

The frequentist coverage rate 
of the Bayes posterior interval for this example  is 
easy to calculate, and is 
shown in Figure \ref{fig:ebcover} for the case that 
$\sigma^2 = \tau^2=1$ and $1-\alpha=0.95$. 
The coverage rate is higher than 95\% if $\mu$ is close to the prior 
mean $\phi$  but approaches zero 
as $\mu$ gets increasingly far away. 
While the coverage rates 
of the empirical Bayes posterior intervals in a multigroup data analysis
will be slightly different, 
it is generally 
the case that the interval for a group 
with a $\mu$-value close to the average 
$\sum_i \mu_i/n$  has a frequentist coverage
above $1-\alpha$, because the interval is shorter
than the UMAU interval and is centered around an estimate that has low bias
for that group. 
Conversely, the interval for a group with a mean far from 
$\sum_i \mu_i/n$ is also narrower than the UMAU 
interval but is centered around an estimate with a high bias, and so the coverage 
rate is lower than $1-\alpha$, and can approach zero as the mean gets 
further away from the other $\mu_i$'s. 
This suggests caution when 
using empirical Bayes posterior intervals:
While these intervals approximately maintain 
a target frequentist coverage rate on average across groups, 
the coverage can be quite poor for outlying groups, which in the examples 
considered include 
students with low test-taking ability, 
or counties with high levels of household radon,
 which are likely the groups of highest concern. 

\section{Information-sharing constant coverage intervals}
If we require our intervals to maintain a 
constant frequentist coverage rate across all groups, 
we could simply use standard frequentist procedures, such as the usual 
$t$-interval in the case that the within-group sampling model is normal.
Such procedures are ``direct'', in that the interval 
for group $i$ does not depend on data from groups other than $i$, 
and so is potentially inefficient. How can ``indirect'' information from 
groups other than $i$ be incorporated into a confidence 
interval procedure for $\mu_i$, while maintaining exact $1-\alpha$ frequentist 
coverage, regardless of the value of $\mu_i$? One answer comes from 
\citet{pratt_1963}, who identified the confidence interval with minimum 
prior risk among those with constant $1-\alpha$ frequentist coverage. 
Specifically, for the case that $Z\sim N(\mu ,1)$, Pratt found the 
interval procedure $C(z)$ that minimizes the prior expected interval width
\begin{equation}
\int   \int  |C(z) | \,  p(z| \mu )dz \,  g(\mu) d\mu 
\end{equation}
subject to the frequentist coverage constraint 
\begin{equation}
\int 1(\mu \in C(z) ) \, p(z|\mu) dz  = 1-\alpha ,\  \ \forall \mu . 
\end{equation}
We refer to this interval procedure as being ``frequentist and 
Bayesian'' (FAB), as it maintains an exact $1-\alpha$ coverage 
rate for every value of $\mu$, while also minimizing a Bayes risk (the prior 
expected interval width). 
The interval can be derived via the duality between a 
$1-\alpha$ confidence procedure and  a collection of 
level-$\alpha$ hypothesis tests: Pratt's FAB interval includes 
the $\mu$-values that are not rejected by level-$\alpha$ hypothesis tests 
that have
optimal prior expected power. 
\citet{yu_hoff_2018} extended Pratt's idea to multigroup data analysis by providing a type of ``indirect'' $t$-interval 
that maintains 
an exact $1-\alpha$ constant coverage rate across groups while 
approximately minimizing the across-group average expected interval width. 
Their interval for a given group $i$ 
is the inversion of a collection of tests for $\mu_i$ that 
maximize prior expected power, where the 
``prior'' distribution is estimated with data from the other groups. 
The frequentist coverage rate is nonparametric, in that it does not 
rely on a correct specification of the 
across-group distribution $G$.

\begin{figure}
\begin{knitrout}
\definecolor{shadecolor}{rgb}{0.969, 0.969, 0.969}\color{fgcolor}

{\centering \includegraphics[width=5in]{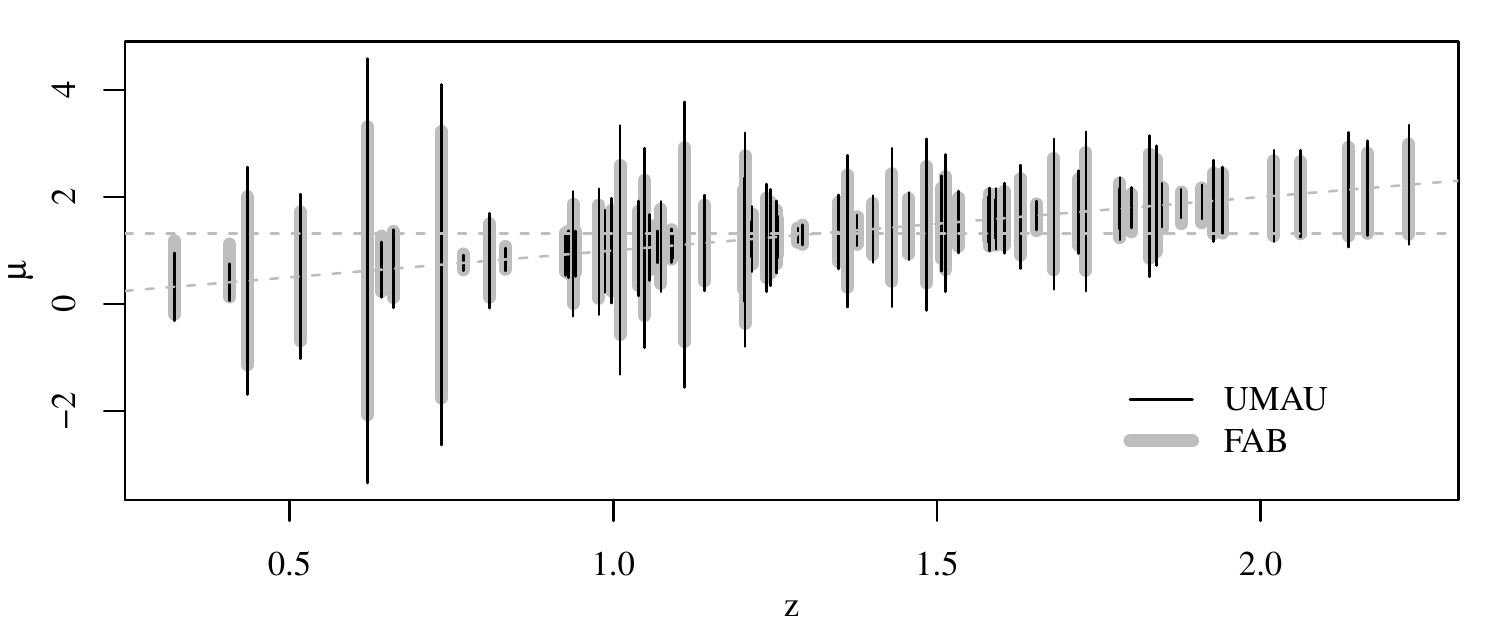} 

}

\end{knitrout}

\caption{95\% FAB and UMAU intervals for the Minnesota radon data.}
\label{fig:fabCI} 
\end{figure}

To illustrate the difference between FAB and standard $t$-intervals in 
a multigroup setting, we recreate the household radon example 
from \citet{yu_hoff_2018}. The data considered include log 
radon levels of 916 households, 
each located in one of 82 counties (these are the counties with 
two or more households in the study). The data and code for 
this example are available via the {\sf R}-package \href{https://cran.r-project.org/web/packages/fabCI/index.html}{\tt fabCI} \citep{hoff_yu_2021}.
We model 
the sample mean in county $i$ as $Z_i \sim N(\mu_i, \sigma^2_i/n_i)$ 
where $\mu_i$ and $\sigma^2_i$ are unknown and 
$n_i$ is the sample size for county $i$. 
After fitting a normal random effects model that presumes 
$\mu_1,\ldots,\mu_n \sim$ i.i.d.\  $N(\phi,\tau^2)$, 
it is ascertained that 
the across-group variance of the $\mu_i$'s is substantially smaller 
than the variability $\sigma^2_i/n_i$ of $Z_i$ for most groups, 
suggesting that across-group information sharing is likely to be
beneficial. County-specific 95\%  FAB intervals are displayed in 
Figure \ref{fig:fabCI}, along with the direct $t$-intervals, 
which are the UMAU intervals for this normal sampling model. 
The FAB intervals are narrower than the UMAU intervals for 77 of the 82 
counties, with UMAU intervals being 30\% wider on average across counties. 
Both procedures have exact $95\%$ frequentist coverage for each 
group (assuming within-group normality of the data), 
regardless of what the true values of $\mu_1,\ldots, \mu_n$ are. 
As discussed above, empirical Bayes posterior intervals lack 
this group-specific 
coverage guarantee.

\section{Summary}
Bayesian methods are often advertised as providing ``a proper accounting of 
uncertainty.''  Too often though, empirical Bayes data analyses ignore 
the uncertainty in the estimation of  the prior distribution. Ignatiadis and Wager highlight this issue and provide useful confidence interval procedures for assessing this uncertainty. As with many empirical Bayes procedures, 
 those of IW focus on across-group inference. 
If interest is instead on group-specific inference, methods with different 
coverage properties may be desired. 

\medskip 

\section*{Acknowledgment} 
I thank Surya Tokdar for discussing this topic with me.

\bibliographystyle{chicago}
\bibliography{ebayesCover}

\end{document}